\documentstyle[sprocl,epsfig,axodraw,rotate]{article}
\def\be{\begin{equation}}
\def\ee{\end{equation}}
\def\bea{\begin{eqnarray}}
\def\eea{\end{eqnarray}}

\begin{document}

\mbox{}
\vspace{-2cm}

\rightline{\vbox{\halign{&#\hfil\cr
WUE-ITP-2004-030\cr
August 2004\cr}}}

\title{Probing Minimal 5D Extensions of the Standard Model}

\author{A.~M\"uck$^1$, A.~Pilaftsis$^2$, and R.~R\"uckl$^1$
 \footnote{Talk presented~at~LCWS2004, Paris, France, April~2004}}

\address{$^1$ Institut f\"ur Theoretische Physik und Astrophysik,
Universit\"at W\"urzburg\\ D-97074 W\"urzburg, Germany}

\address{$^2$ Department of Physics and Astronomy, University of Manchester,\\
         Manchester M13 9PL, United Kingdom}


\maketitle\abstracts{
We analyze non-universal 5D   standard model extension, where  some or
all of the gauge and Higgs fields propagate in a flat extra dimension,
while   all  other degrees  of freedom  are  localized  on a $S^1/Z_2$
orbifold  brane. From     LEP  data, model-dependent  bounds  on   the
compactification scale $M$ between 4 and 6~TeV are derived. We analyze
the  correlations   between   $M$  and  the  SM    Higgs   mass~$m_H$.
Investigating the  prospects  at an  $e^+e^-$ linear collider  such as
TESLA, we  show that the so-called  GigaZ option has  the potential to
improve  the LEP bounds by about  a factor 2.   At  the center of mass
energy    of 800~GeV    and      with an   integrated   luminosity  of
$10^3$~fb$^{-1}$,     linear       collider  experiments   can   probe
compactification scales up to 20--30~TeV  and beyond, depending on the
systematic errors.  }

The fascinating possibility that our world  may realize more than four
dimensions    has been   a  central       theme  of   the last     ten
years~\cite{ADD,CC,DDG,FF}.  Considering higher-dimensional extensions
of the  standard model, the lower  limit on the compactification scale
$M$  of a single  extra dimension is roughly  300~GeV in the so-called
universal scenario, in which all standard model  fields live in the 5D
bulk.   For non-universal models,  i.e. if some of   the SM fields, in
particular the fermions, are confined   to the familiar  4-dimensional
world,   $M$    is   constrained much      more  strongly,  namely  $M
\stackrel{>}{{}_\sim}  4$~TeV~\cite{MPR2}.  This  order  of  magnitude
increase of the bound is due  to the tree-level  coupling of single KK
excitations  to light  SM  modes   on  a  brane  which induce  contact
interactions at  low energies. Moreover,  a Higgs field  confined to a
brane implies shifts in gauge-boson couplings and masses.

Here, we distinguish three non-universal five-dimensional scenarios on
an $S^1/Z_2$ orbifold~\cite{MPR2,MPR}:  (i)~all gauge fields propagate
in the bulk of the  extra  dimension (bulk-bulk model); (ii)~only  the
SU(2)$_L$ gauge bosons are bulk fields, while the U(1)$_Y$ gauge field
is  confined  to a  brane (bulk-brane model);  (iii)~only the U(1)$_Y$
boson propagates in the bulk (brane-bulk model). While the Higgs field
has to be confined to the brane in the latter  two models, it may also
propagate in the extra dimension in the bulk-bulk  model.  As has been
shown in~\cite{MPR}, the above 5D models can be consistently quantized
using appropriate  5D gauge-fixing conditions that  lead, after the KK
reduction, to the known class of $R_\xi$ gauges.

\begin{figure}[t]
\begin{center}
\includegraphics[width=11.9cm]{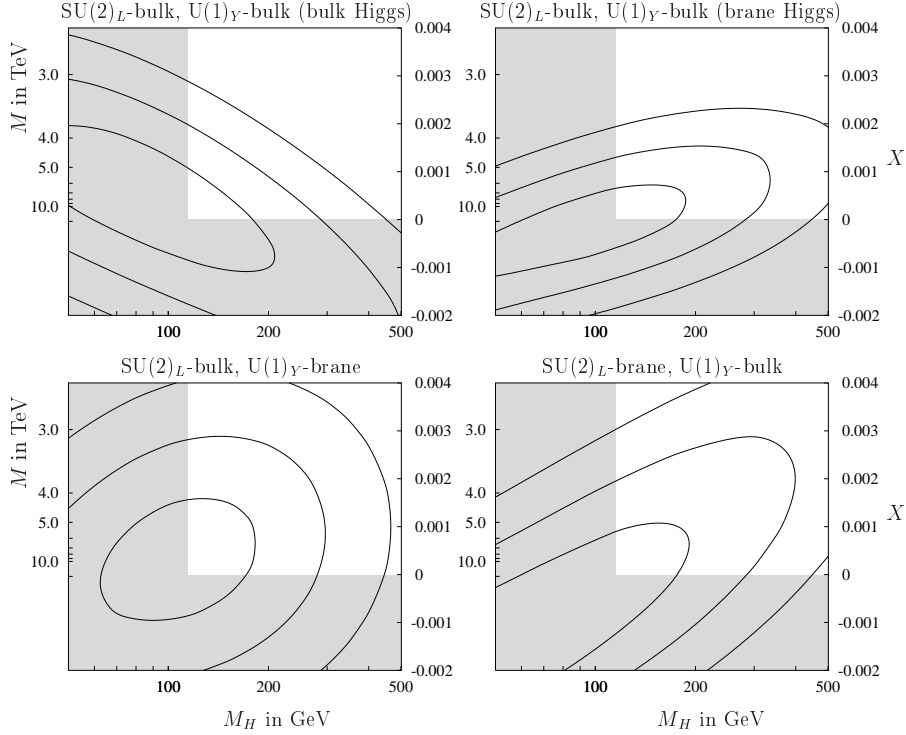}
\end{center}
\caption{\label{contoursprecobs} Contours of 
$\Delta \chi^2  = 1,\, 4,\, 9$ derived
from  multi-parameter  fits  to electroweak precision data. The shaded
regions of the parameter space  correspond  to $m_H < 114$~GeV  and/or
$M^2 < 0$.}
\end{figure}

Concerning  the  phenomenology of such  extra-dimensional extension of
the standard model, the shifts in gauge-boson masses and couplings due
to a  Higgs boson which is  confined to a brane  are  of the  order of
$X=\pi^2 m_Z^2/3M^2$.  Moreover,   input parameters  like  the   Fermi
constant $G_F$ and the $Z$  mass $m_Z$ have  to be reinterpreted.  For
example, the KK modes of  the $W$ boson contribute  to muon decay  and
thus enter  the determination of  $G_F$. These shifts are particularly
important for  precision  physics at the  $Z$  pole, where  virtual KK
exchange is negligible. To      confront the different  models    with
experiment, we calculate the tree-level shift $\Delta^{\rm 5DSM}_{\cal
O}$   for a  given observable due   to  the extra dimension.  Applying
$\Delta^{\rm  5DSM}_{\cal O}$     to the   radiatively   corrected  SM
prediction  ${\cal   O}^{\rm SM}$  to   obtain  the higher-dimensional
prediction  ${\cal  O}^{\rm 5DSM} \! =\!  {\cal  O}^{\rm SM} \big( 1 +
\Delta^{\rm 5DSM}_{\cal O} \big)$,  we perform a $\chi^2$ analysis for
a large number of precision observables.  Using ZFITTER~\cite{ZFITTER}
to  calculate  ${\cal O}^{\rm  SM}$,   a  multiparameter fit  of   the
compactification scale $M$  along   with  the standard  model    input
parameters yields bounds on $M$ between 3~TeV for the brane-bulk model
and 4~TeV for the bulk-bulk model with a brane Higgs at the $2 \sigma$
confidence level.  The correlation between  the SM  Higgs mass and the
compactification    scale~\cite{RW}           is        displayed   in
Fig.~\ref{contoursprecobs}.  One sees that in some  of  the models the
existence  of an extra  dimension favors  a heavier  Higgs boson.  The
effect is  most pronounced in the bulk-bulk  model with  a brane Higgs
and in the brane-bulk model. Quantitatively, for $M = 5$~TeV, the best
fit   values   increase from $m_H  =  100^{+70}_{-40}$~GeV   to $m_H =
170^{+105}_{-60}$~GeV and  $m_H = 155^{+105}_{-60}$~GeV, respectively.
If the compactification scale is  included in the multi-parameter fit,
the 2$\sigma$ upper bound on $m_H$ is relaxed  from 280~GeV to 330 and
400~GeV,  respectively. Note, that   the data set~\cite{MPR2} used for
this analysis is not exactly the same as the one used for the familiar
blue-band plot and that the recent change in the top mass is not taken
into account.

\begin{figure}[t]
\begin{center}
\includegraphics[width=11.5cm]{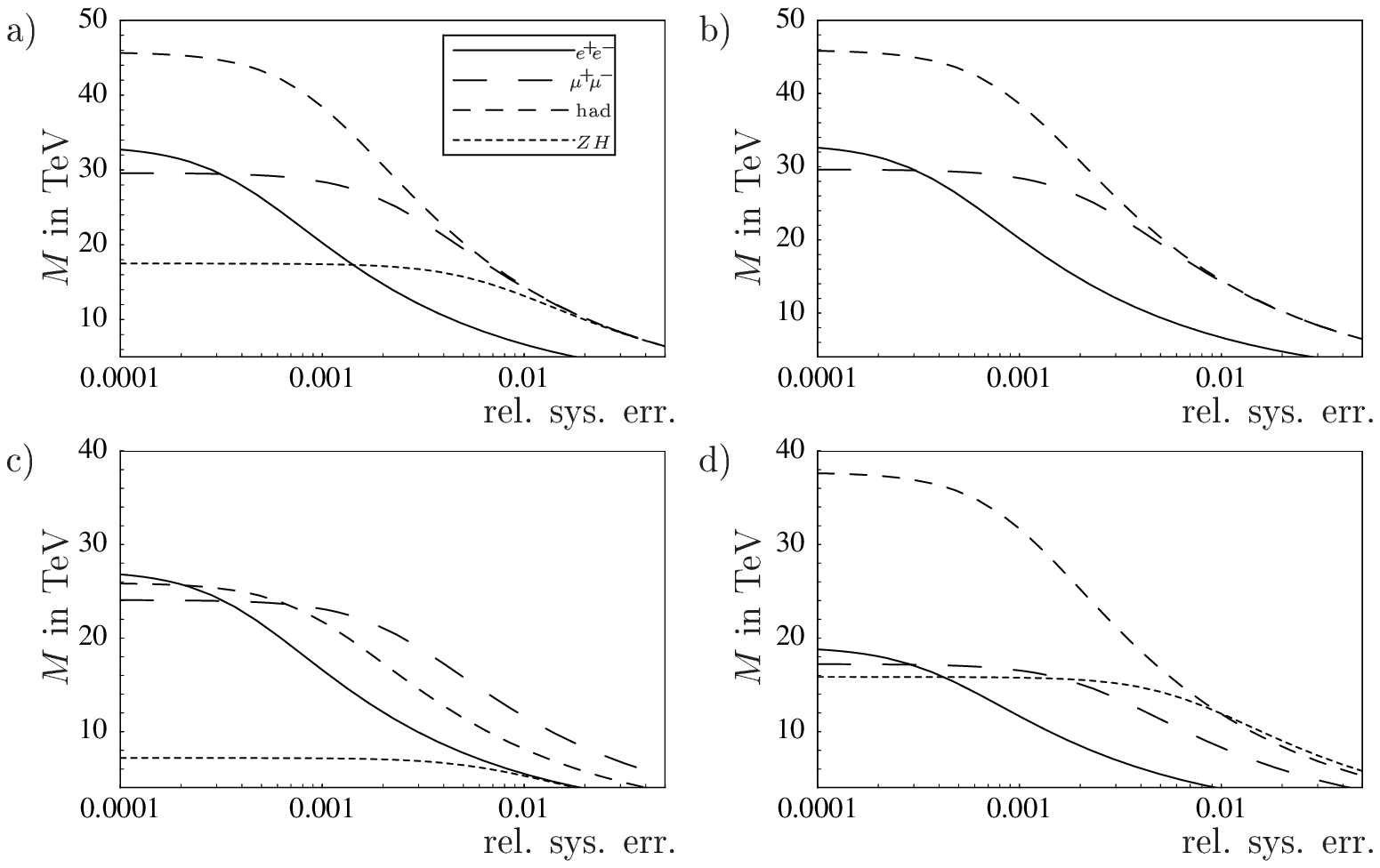}
\end{center}
\caption{\label{TESLAvariables} Expected sensitivity to the 
compactification scale  $M$ as a function  of  the relative systematic
error for Bhabha scattering,  muon-pair production, hadron production,
and   Higgsstrahlung  at $\sqrt{s} = 800$~GeV    and for an integrated
luminosity of  1000~fb$^{-1}$: (a)  bulk-bulk model with  brane Higgs,
(b)  bulk-bulk model with bulk   Higgs, (c) brane-bulk  model, and (d)
bulk-brane model.}
\end{figure}

Already at LEP2 energies, the exchange of heavy KK modes dominates the
higher-dimensional  shifts  for fermion-pair production. Investigating
the cross section for Bhabha scattering and the production of muon and
tau  pairs,  hadrons, and  $W$ pairs,   the  compactification scale is
constraint to be larger than 4~TeV for the brane  bulk model and 6~TeV
for the bulk-bulk model with a brane Higgs boson. Muon-pair, tau-pair,
and  hadron production have been  again  analyzed in a multi-parameter
fit. The correlations of $M$ with the Higgs mass only change little.

At  a future $e^+e^-$  linear collider,  the GigaZ  option  allows for
tests of the higher-dimensional shifts  in  masses and couplings  with
increased statistics.  In  particular, the precise  measurement of the
left-right asymmetry  $A_{\mathrm{LR}}$ using  polarized beams  can be
utilized to explore the compactification  scale. GigaZ is sensitive to
a compactification scale $M$  from 7~TeV for  the bulk-brane  model to
about 12~TeV for the other models considered. At $\sqrt{s} = 800$~GeV,
the  contact-like interactions due to  the  exchange of heavy KK modes
completely dominate    the  higher-dimensional   shifts   of     cross
sections. For an integrated luminosity of 1000~fb$^{-1}$ the discovery
potential  will crucially depend on the  control of systematic errors,
as  shown  in     Fig.~\ref{TESLAvariables}   for  different    search
channels. For  a systematic uncertainty  of 1\% in each search channel
and combining all  data, one  will be  able  to reach compactification
scales in the  range 15-20~TeV.  For systematic uncertainties  smaller
than the  statistical uncertainties,   the  sensitivity limit    would
saturate in  the range $M =$ 35-50~TeV.   Moreover, for a sufficiently
low  compactification   scale,  $M   \stackrel{<}{{}_\sim}    10$~TeV,
Higgsstrahlung and angular distributions of 2-fermion final states can
be used to discriminate   between different 5D models. In  particular,
Higgsstrahlung  is useful  for  distinguishing brane  from  bulk Higgs
bosons.

\smallskip
\noindent
This work  was supported by  the  Bundesministerium f\"ur Bildung  und
Forschung (BMBF, Bonn,  Germany) under the  contract number 05HT4WWA2,
and \linebreak PPARC grant number PPA/G/O/2000/00461.

\end{document}